# Carbon ene-yne graphyne monolayer as an outstanding anode material for Li/Na ion batteries

Meysam Makaremi,[1] Bohayra Mortazavi,[2] and Chandra Veer Singh*,[1,3]

[1]Department of Materials Science and Engineering, University of Toronto, 184 College Street, Suite 140, Toronto, ON M5S 3E4, Canada.
[2]Institute of Structural Mechanics, Bauhaus-Universität Weimar, Marienstr. 15, D-99423 Weimar, Germany.
[3]Department of Mechanical and Industrial Engineering, University of Toronto, 5 King's College Road, Toronto M5S 3G8, Canada.

**Abstract**

Recently "carbon ene-yne" (CEY), a novel full carbon two-dimensional (2D) material was successfully synthesized by the solvent-phase reaction. Motivated by this experimental effort, we conducted extensive first-principles density functional theory simulations to explore the application prospects of CEY as an anode material for Mg, Na and Li-ion batteries. To simulate the ionic intercalation process in an anode electrode, the adatoms coverage was gradually increased. We then employed Bader charge analysis to evaluate the charge transfer between the adatoms and the CEY nanosheet and finally to report the theoretical storage capacity of the anode. We particularly studied the evolution of adsorption energy, electronic density of states and open-circuit voltage with respect to adatoms coverage. The diffusion of an adatom over the CEY surface was also investigated by using the nudged elastic band method. Remarkably, our results suggest CEY as a promising anode material containing the highest theoretical charge capacities among 2D materials studied so far, with ultrahigh capacities of 2,680 mAh/g and 1,788 mAh/g for Li and Na-ion batteries, respectively. The provided insight by this study highlights the CEY as a novel full carbon material with properties, highly desirable for the application as anode material in the next generation rechargeable ion batteries.

**1. Introduction**

After the first successful fabrication of graphene from graphite by using mechanical exfoliation method in 2004 [1–4], two-dimensional (2D) materials have emerged as a new class of materials with intriguing characteristics [5,6]. Due to subsequent evolution, currently there exist a wide variety of 2D materials that can be synthesized at large scale and high quality. The 2D materials made only from carbon atoms also include a wide range of structures, as they were theorized by Baughman *et. al* [7] in 1987. Many 2D carbon allotropes are in the form of graphyne [7], which includes sp and $sp^2$ hybrid bonded carbon atoms arranged in a crystal lattice. Graphyne structures involve lattices of benzene rings connected by acetylene bonds [8]. Despite of tremendous successes in the synthesis of various 2D materials, the isolation of fabricated nanosheets from the substrate have remained among the most challenging issues [9–14]. Nonetheless, most recently a remarkable experimental advance was accomplished and carbon ene-yne (CEY), a full carbon 2D material was synthesized with different thicknesses from tetraethynylethene by the solvent-phase reaction on the cupper substrate [15]. These nanosheets were then successfully separated from the substrate and tested for the Li-storage, confirming that CEY nanomembranes can be fabricated in free-standing form, which is a crucial requirement for the application of a 2D material as an electrode in rechargeable batteries. Notably, in this first experimental work, the CEY nanosheets have been confirmed to yield very promising performances for the Li-ion storage [15]. The atomic structure of CEY is similar to graphyne crystal, but lacking carbon benzene rings. Worthy to mention that the formation of CEY was first predicted theoretically by Baughman *et. al* [7] (CEY was introduced as 14, 14, 14-graphyne). The successful synthesis of 2D CEY consequently raise the importance of experimental and theoretical studies to shed light on its intrinsic properties and its performance for various applications. In the recent theoretical investigation [16], it was found that CEY can yield attractive electronic, thermal conduction and optical properties, very promising for a wide variety of device applications.

In recent years, increasing technical development of rechargeable metallic ion batteries has been playing an essential role in the progress in electronic, communication devices and automobile industries [17,18]. Currently, graphite is the mostly employed anode materials in Li-ion batteries. However, the charge capacity of graphite is low (372 mAh/g [19]) and to improve the efficiency of rechargeable batteries, more advanced anode materials with higher capacities are needed to replace the graphite. Such that during the last decade tremendous experimental efforts have been devoted to find alternative anode materials. In this regard, silicon anodes have been taken into consideration, yielding specific capacity of 4200 mAh/g [20]. Nonetheless, because of their massive volume changes during ionic intercalations which leads to the structural degradation, the commercialization of these anode materials has been highly prohibited [21,22]. On the other hand, 2D materials and their hybrid structures owing to their large surface area and remarkable thermal and mechanical stabilities are currently considered as promising solutions to reach high efficient rechargeable batteries. Previous theoretical studies have confirmed that 2D materials and their heterostructures can yield remarkably high charge capacities [23–25] and low diffusion energy barriers [26–30].

Motivated by the successful synthesis of CEY involving structural stability and complete isolation from the substrate, we conducted extensive density functional theory (DFT) calculations to probe the application of 2D CEY as an anode material for rechargeable Na, Mg, and Li-ion batteries. Since the charge capacity of an anode material for a particular metal atom storage directly correlates to the adsorption energy, we studied the evolution of adsorption energy by increasing the adatoms coverage. In addition, we investigated the open-circuit voltage profiles and electronic characteristics of the CEY nanosheet covered with different concentrations of metal adatoms. We employed the nudged elastic band (NEB) method to probe the diffusion path and corresponding energy barrier for considered adatoms over the CEY surface. The thermal and mechanical stability of saturated CEY films were also assessed by the *ab-initio* molecular dynamics (AIMD) simulations. We finally report the theoretical specific storage capacity of 2D CEY for different types of adatoms, according to the Bader charge analysis. Our extensive first-principles results propose the 2D CEY as an outstanding anode material for rechargeable ion batteries with ultrahigh specific charge capacities outperforming all other 2D materials.

## 2. Computational Details

Spin polarized density functional theory (DFT) simulations were carried out via using generalized gradient approximation (GGA) with the Perdew–Burke–Ernzerhof (PBE) functionals [31] and projector augmented-wave (PAW) potentials [32], as implemented in the Vienna *Ab-initio* Simulation Package (VASP) [33] framework. The kinetic energy cutoff, electron self-consistent convergence threshold and Hellmann–Feynman force convergence criterion were chosen to be 500 eV, $1\times10^{-6}$ eV and $1\times10^{-3}$ eV/Å, respectively. The tetrahedron method with Blöchl corrections, and the Monkhorst-Pack scheme [34] with a mesh grid of 15x15x1were considered for smearing, and integrating the Brillouin zone, accordingly. As discussed in the literature, the GGA underestimates the binding energies [35–37]. However, in this work we used GGA+VDW, using the semiempirical correction of Grimme [38], which modifies the binding energies., the GGA+VDW exchange correlation gives better agreement with experimental cohesive energies as well as binding energies for several systems. Furthermore, Heyd-Scuseria-Ernzerhof (HSE06) functional [39] calculations were also used to accurately evaluate the electronic density of states.

As shown in Figure 1, the CEY unitcell consists of 20 carbon atoms and involves a monoclinic structure with lattice constants a = 11.26 Å and b = 9.74 Å. The CEY atomic configuration and lattice parameters are listed in Tables S1 and S2. A supercell composed of 1x2 units including 40 carbon atoms, and a vacuum space of 20 Å were considered. To find the most stable adsorption configurations, ten possible binding sites were considered for each adatom as illustrated in Fig. 1b. Next, the adatoms were added to the surface at random locations (4 atoms each time) until reaching to the maximum coverage of the surface leading to the optimal capacity. To evaluate the thermal stability of the final structure including the optimal capacity, we also carried out *ab-initio* molecular dynamics (AIMD) simulations in canonical ensemble (NVT) including a time step of 1 fs for 10000 steps at 300 K.

The adsorption energy resulting from the interaction of the monolayer with a adatoms is calculated by using the following formula,

$$E_{\text{av-ad}} = \frac{(E_{\text{CEY}+M} - n \times E_M - E_{\text{CEY}})}{n}, \qquad (1)$$

where $E_{\text{CEY}+M}$, $E_{\text{CEY}}$, $E_M$, and n are the total interaction energy, the energy of the bare monolayer, the energy of the single adatom, and the number of adatoms; respectively. Negative values of $E_{\text{ad}}$ indicate the binding between the monolayer and adatom.

To study the local charge density, we used the Bader analysis approach [40]. The total charge difference can be calculated as,

$$\Delta\rho = \rho_{\text{CEY}+M} - \rho_{\text{CEY}} - \rho_M, \qquad (2)$$

here, $\rho_{\text{CEY}+M}$, $\rho_{\text{CEY}}$, and $\rho_M$ include the electron charge densities of CEY/M, CEY, and the adatom, respectively.

Storage capacity is a key factor determining the performance of an ion battery. It can be calculated from Faraday equation,

$$q = 1000 \, F \, z \, \frac{n_{max}}{M_{CEY}}, \qquad (3)$$

where F, z, $n_{max}$, and $M_{CEY}$ are the Faraday constant, adatom valence number, optimal concentration of adsorbed adatoms, and the molecular mass of CEY monolayer. To ensure about the performance of the battery, another factor, open-circuit voltage ($V$) [41], need to be probed for the coverage ratio of $x_1 \leq x \leq x_2$ *i.e.* the average number of adsorbed adatoms to the number of bare monolayer atoms, and is described as,

$$V \approx \frac{\left((x_2 - x_1)E_M - (E_{M_{x_2}C} - E_{M_{x_1}C})\right)}{(x_2 - x_1)e}, \qquad (4)$$

here, $E_{M_{x_1}C}$, $E_{M_{x_2}C}$ are the total energies for the surface coverages $x_1$ and $x_2$, respectively, and e is the electron charge.

## 3. Results and Discussion

The most favorable adsorption sites involving the lowest adsorption energies for Li, Na, and Mg adatoms are shown in Figure 2. Although for all adatoms the adsorption happens at the same place with respect to the planar directions x and y, for the case of Li and Na the adatom moves to the inside of the monolayer (the distance to the surface in the *z* direction is equal to zero). By using Eqn. (1) for a single adatom, the optimal adsorption energies are computed as -1.098 eV, -1.226 eV, and 1.262 eV for Li, Na and Mg adatoms, respectively, which show that while Li and Na bind strongly to the CEY surface, the binding of Mg is not energetically favorable. Moreover, we calculated the differential charge density via Eqn. (2). Figure 3 suggests that Li and Na adsorptions result in a complete charge transfer between the adatom and the monolayer due to the acceptance of all valence electrons of adatom, whereas for the Mg adatom the charge transfer is partial. Therefore, CEY is not a suitable candidate electrode material for Mg batteries; henceforth we only consider the electronic structure properties of the CEY nanosheet interacting with Li and Na, and investigate the feasibility of the nanosheet to be used as a potential anode for ion battery applications.

After finding the most stable sites on the monolayer, the surface was gradually covered with adatoms until the maximum charge capacity was reached, the point after which there was no more charge transfer from adatoms to the surface. Using the Bader analysis, we were able to find the saturation limit of the charge transfer, equivalent to the maximal charge capacity of the structure, occurring at a certain coverage at which the anode surface cannot accept additional charges due to further increasing the number of adatoms. As depicted in Figure 4, more Li adatoms can be adsorbed on the CEY surface, compared to Na foreign atoms; however, the number of adsorbed Na adatoms is still considerable. Our simulations show that CEY can adsorb up to 24 Li or 16 Na adatoms per unitcell leading to storage charge capacities of 2,680 and 1,788 mAh/g, respectively. These results confirm the ultrahigh charge capacity of CEY for Li or Na ions storage. We remind the theoretical charge capacity of graphite for Li ion anode application is 372 mAh/g [42]. Moreover, the maximum charge capacity of a graphyne structure for Li ions was predicted to be 1117 mAh/g [43]. To date, the flat borophene [44] films have been theoretically predicted [23] to yield the highest capacities among the all other 2D materials with the maximum charge capacities of 1980 mAh/g and 1640 mAh/g for the Li and Na ions storages, respectively. As it is clear, CEY distinctly outperform not only the borophene films but also the other 2D materials with respect to the charge capacity.

To ensure about the stability of the structure, AIMD simulations were carried out at 300 K for the systems involving the optimal capacity (see Figure 4, last column). The variations of the energy and the temperature as a function of the simulation time are shown in Figures S1 and S2. Our AIMD simulations confirm that the upon the adatoms adsorptions the CEY films remain intact and such that they are thermally stable for the application as an anode material.

To further understand the nature of adatom binding, the electronic density of states (DOS) for bare CEY, and the monolayer interacting with various numbers of Li or Na atoms were evaluated by using PBE and HSE methods, and the acquired results are illustrated in Figures 5 and 6, respectively. The bare monolayer presents semiconducting behavior including a tiny PBE band gap of 0.05 eV, and a larger HSE energy gap of 0.54 eV consistent with those reported in the recent theoretical study [16] ($E_{g\_PBE}$ = 0.04 eV and $E_{g\_HSE06}$ = 0.54 eV). As confirmed by the both PBE and HSE06 methods, upon the adsorption of Li or Na adatoms, the structures yield metallic electronic character which is a highly promising feature for the application of a material as an electrode in rechargeable metal-ion batteries. It should be noted that internal electronic resistances and corresponding voltage drop and ohmic heats generated during the battery operation are directly proportional to the electronic conductivity of the electrode materials. In this case, the electronic conductivity of CEY is therefore highly desirable for its application as an anode material.

The average adsorption energy, and open circuit voltage as a function of the coverage ratio is illustrated in Figure 8. For both of the elements, the average adsorption energy declines with respect to x due to the increased interatomic repulsion of adatoms. The initial adsorption energy for Na is slightly higher than that for Li; although, for the higher ratios the energy of Li adsorption is more pronounced and the monolayer can accept a larger number of adatoms.

Figure 8 illustrates that the adsorptions of both adatoms lead to positive voltage values during the whole coverage spectrum showing that the adsorptions are favorable, since a negative open circuit voltage indicates that adatoms tend to form metallic states instead of binding to the surface. The voltage for low Na coverage (1.08 V) is higher than the one for Li (0.75 V); whereas, the whole voltage range for Li (0.23-0.75 V) consists of smaller changes with respect to the coverage ratio compared to Na (0.1-1.07 V). It is worthwhile to note that a voltage spectrum of 0.1-1 V is favored for an anode electrode [28], highly confirming the applicability of the CEY monolayer for anodic applications. For instance the potential range for brophene and $TiO_2$ anodes involves 0.5-1.8 V, and 1.5-1.8 V; respectively [23,45].

Nudged-elastic band calculations were performed to study the diffusion behavior of adatoms on the CEY surface. As depicted in Fig. 8a, two diffusion paths called horizontal and vertical pathways were considered for each adatom. Energy barriers include 0.60 eV, 0.58 eV, 0.58 eV, and 0.56 eV for horizontal Li, vertical Li, horizontal Na, and vertical Na diffusions, respectively. For comparison, we note that the Li ions diffusion energy barrier on $Ti_3C_2$ MXene is ~0.70 eV [46], on phosphorene it is 0.13–0.76 eV [47], on graphene it is ~0.37 eV [48], on silicene it is 0.230 eV [49], on germanene or stanene it is 0.25 eV [50] and on the flat borophene is 0.69 eV [23]. As it is clear, the predicted energy barrier for the Li adatom diffusion over the CEY is close to the upper bond of the barrier spectrum for other 2D materials. This observation suggests the slight limitation of CEY in achieving fast charging or discharging rates as compared with other 2D materials. Although, the indicated diffusion barrier is still within the energy barrier range for commercially used anode electrodes based on $TiO_2$ including a spectrum of 0.35-0.65 eV [45,51,52].

## 4. Conclusions

We performed extensive first principle PBE and HSE DFT simulations to study the potential application of carbon ene-yne (CEY), a newly synthesized 2D full carbon structure, as a new anode material for rechargeable Li, Na, and Mg ion batteries. Various anodic aspects including the electronic density of states, differential charge transfer, adsorption energy, open circuit voltage, storage capacity, and adatoms diffusion were investigated.

The CEY monolayer presents strong binding energies upon the adsorption of Li and Na adatoms, while it repels Mg adatoms; therefore, this study suggests that CEY cannot serve as an anode material for Mg-ion batteries. Pristine CEY shows semiconducting properties; and the adsorption of Li or Na adatoms induces the metallic behavior to the nanosheet, which is highly demanded for an anode material. In addition, open circuit voltage and atomic diffusion calculations confirms the feasibility of CEY as a suitable anode material.

This novel carbon 2D structure illustrates ultrahigh capacities of 2,680 mAh/g and 1,788 mAh/g for Li and Na-ion batteries, respectively, which distinctly outperform other 2D materials. Our first-principles modelling results propose the CEY graphyne as an outstanding 2D material for the application as anode materials, owing to its ultrahigh charge capacity, good electronic conductivity, desirable open-circuit voltage and acceptable thermal stability. We therefore hope that this study can open a new horizon to develop a new generation of anode materials with ultrahigh capacity, and to improve the efficiency of the current rechargeable ion batteries.

AUTHOR INFORMATION

**Corresponding Author**

* chandraveer.singh@utoronto.ca

ACKNOWLEDGMENT

CVS and MM gratefully acknowledge their financial support in parts by Natural Sciences and Engineering Council of Canada (NSERC), University of Toronto, Connaught Global Challenge Award, and Hart Professorship. The computations were carried out through Compute Canada facilities, particularly SciNet and Calcul-Quebec. BM acknowledges support from European Research Council for COMBAT project (Grant number 615132).

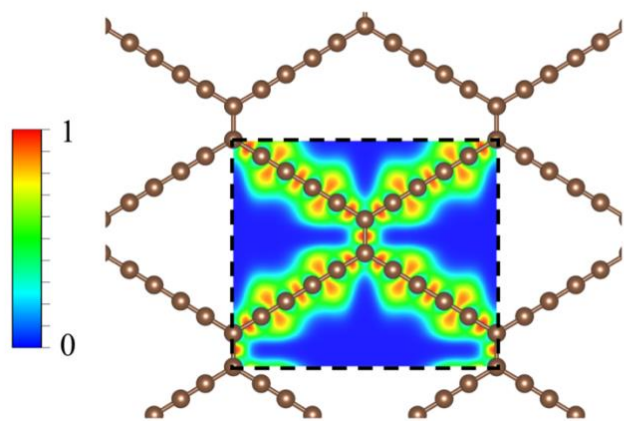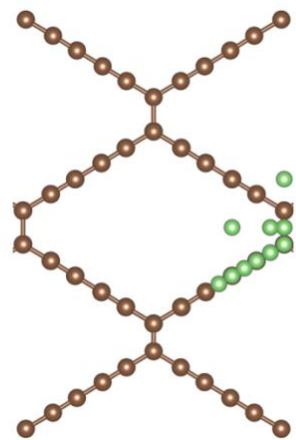

Figure 1. a) The CEY structure and unitcell (black dashed line) including carbon atoms. The contours illustrate electron localization function (ELF), which has a value between 0 and 1, where 1 corresponds to perfect localization. b) CEY adatom adsorption sites. Color coding includes brown and green for carbon and adatom, respectively.

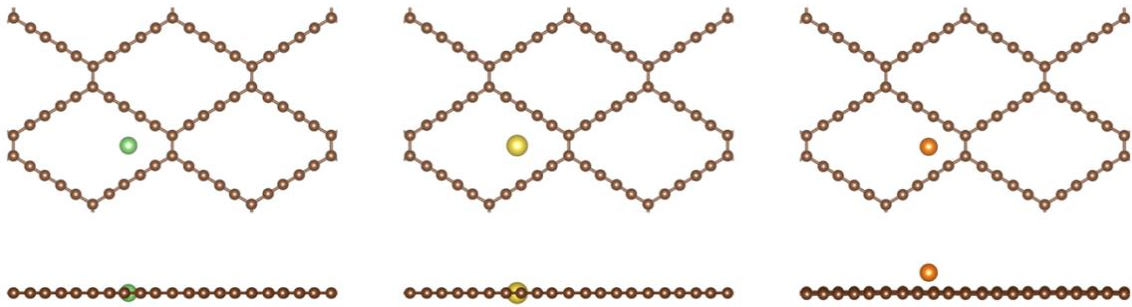

Figure 2. Adsorption configurations of different adatoms: a) Li (green), b) Na (yellow), and c) Mg (orange), over the CEY monolayer.

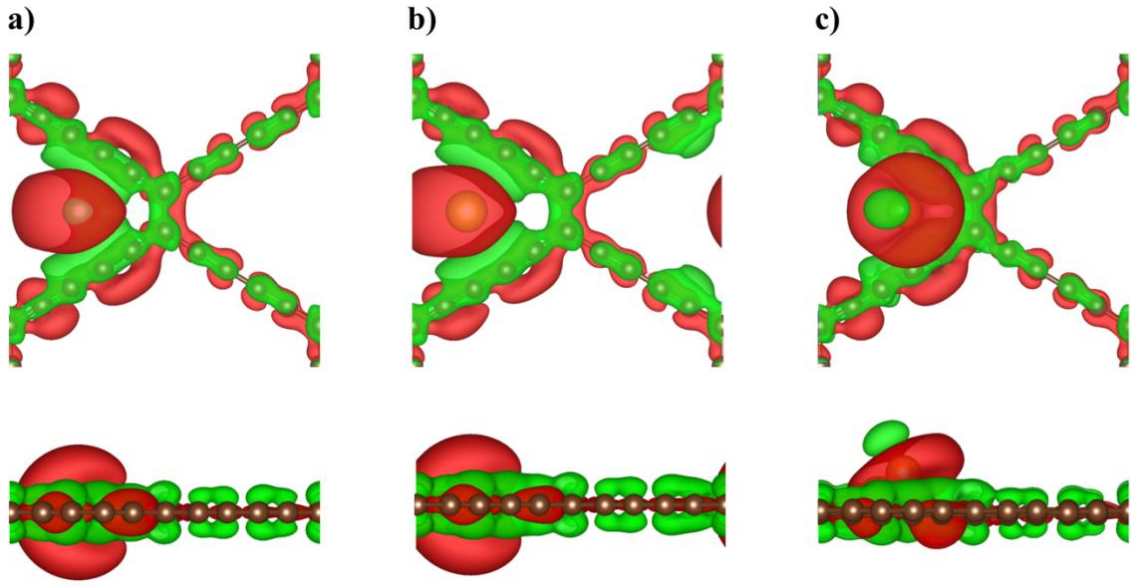

Figure 3. Differential charge density due to the adsorption of different adatoms: a) Li, b) Na, and c) Mg, on the CEY surface. Color coding consists of red for charge gain and green for charge loss.

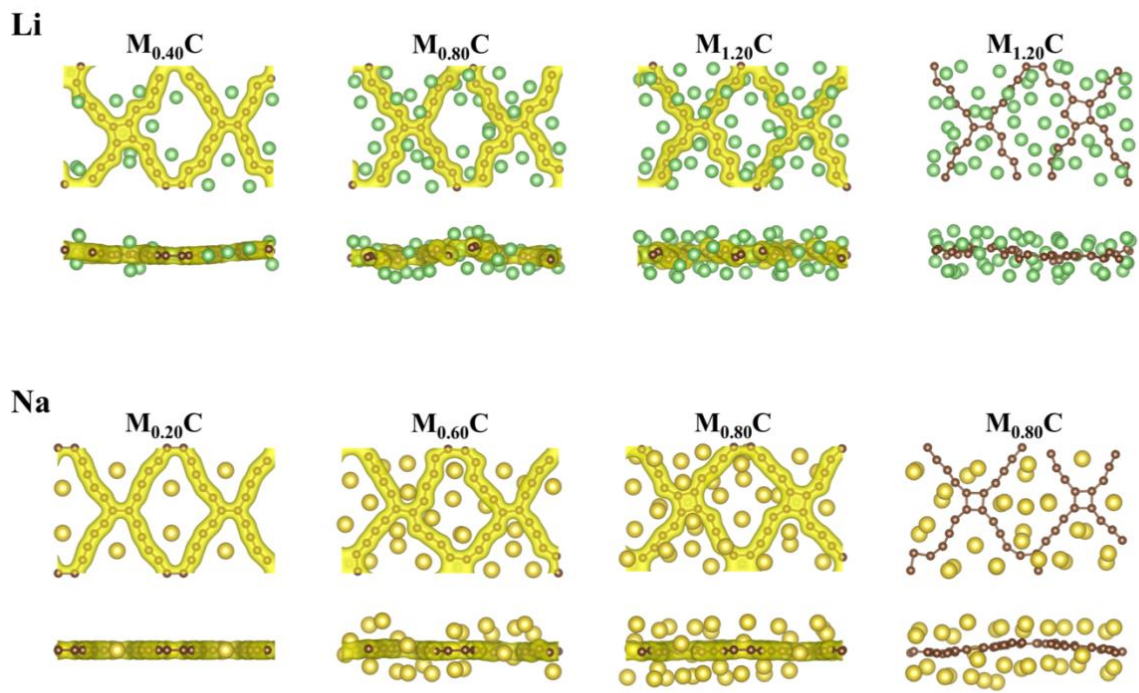

Figure 4. Top and side views of Li and Na atomic adsorption on CEY containing different numbers of adatoms. The third column illustrate the configuration which results in the maximal possible capacity for the CEY monolayer, and the fourth column shows the AIMD results for corresponding configurations at 300 K. Brown, green, and yellow balls represent C, Li, and Na atoms, respectively. The yellow contour indicates the charge density of the CEY nanosheet.

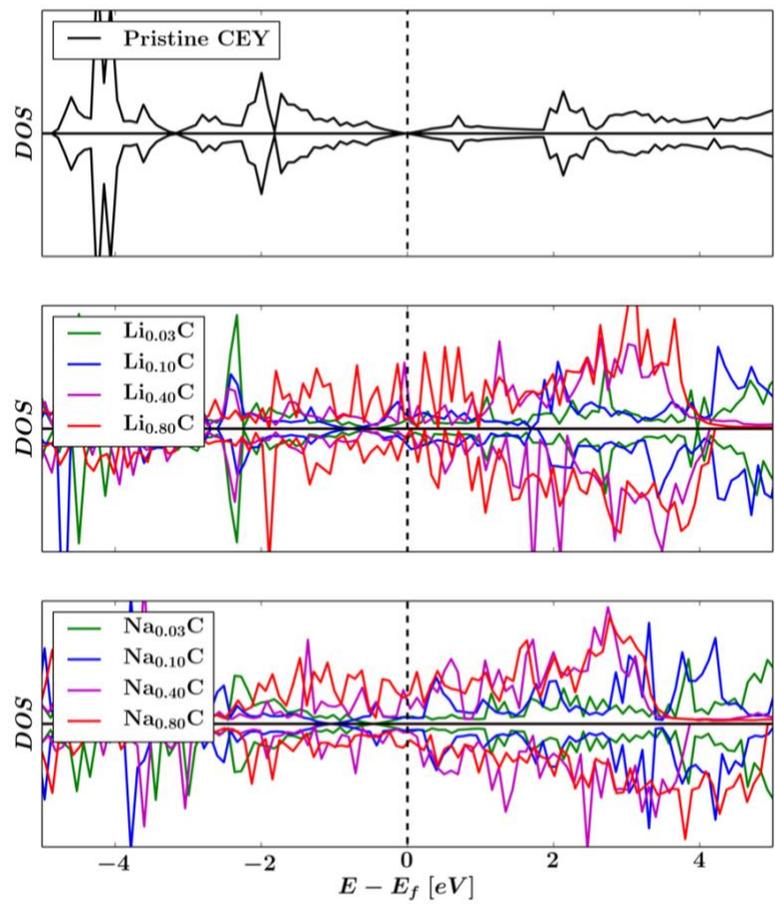

Figure 5. PBE Spin polarized density of states for the pristine CEY monolayer, and the monolayer covered with various concentrations of Li or Na adatoms. Black dashed line represents the Fermi-level.

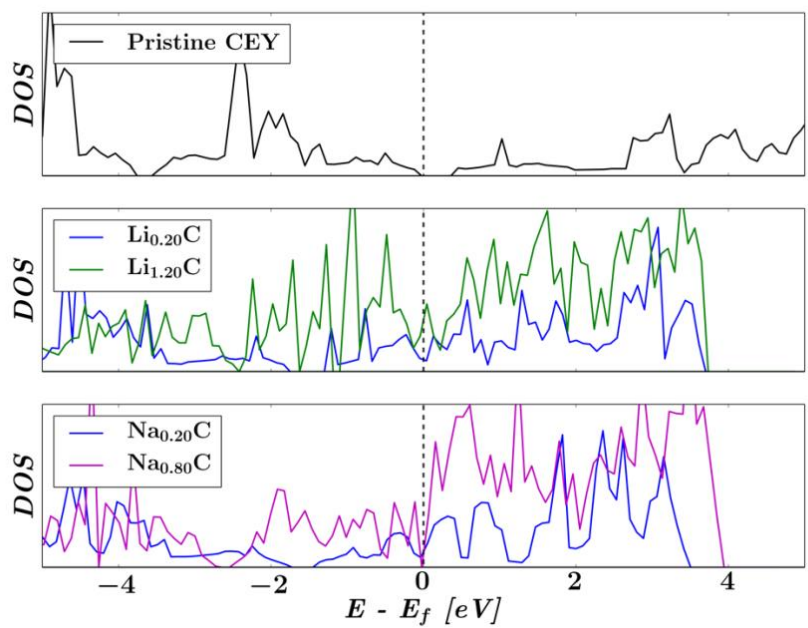

Figure 6. Spin polarized density of states for the pristine CEY monolayer, and the monolayer covered with various concentrations of Li or Na adatoms. Black dashed line represents the Fermi-level.

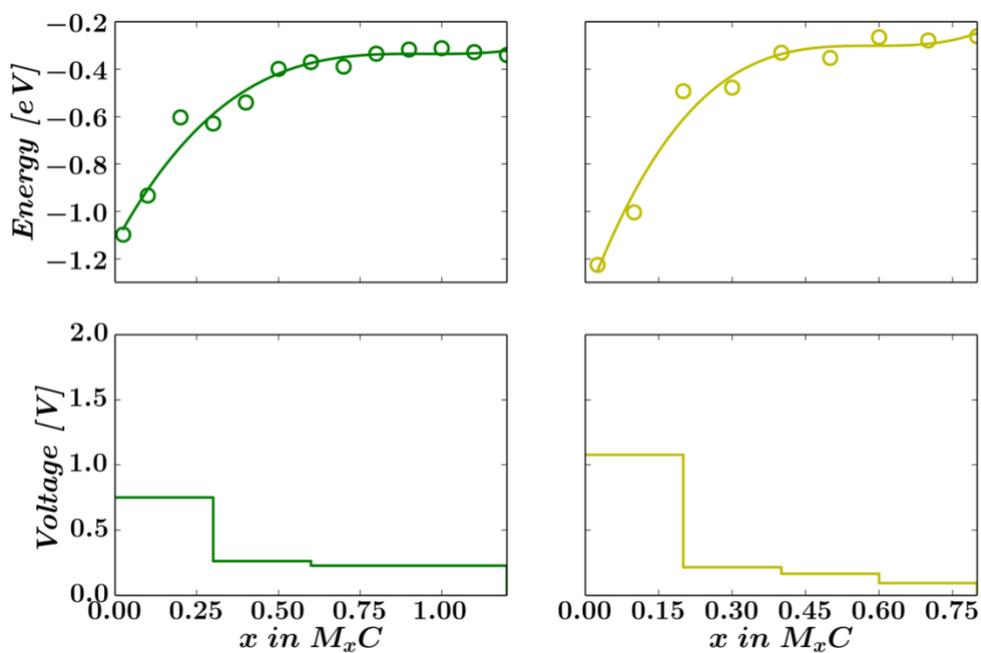

Figure 7. Average adsorption energy and open-circuit voltage with respect to the adatom coverage (x) for Li and Na intercalation in CEY. Color coding involves green and yellow for the Li and Na coverage, respectively.

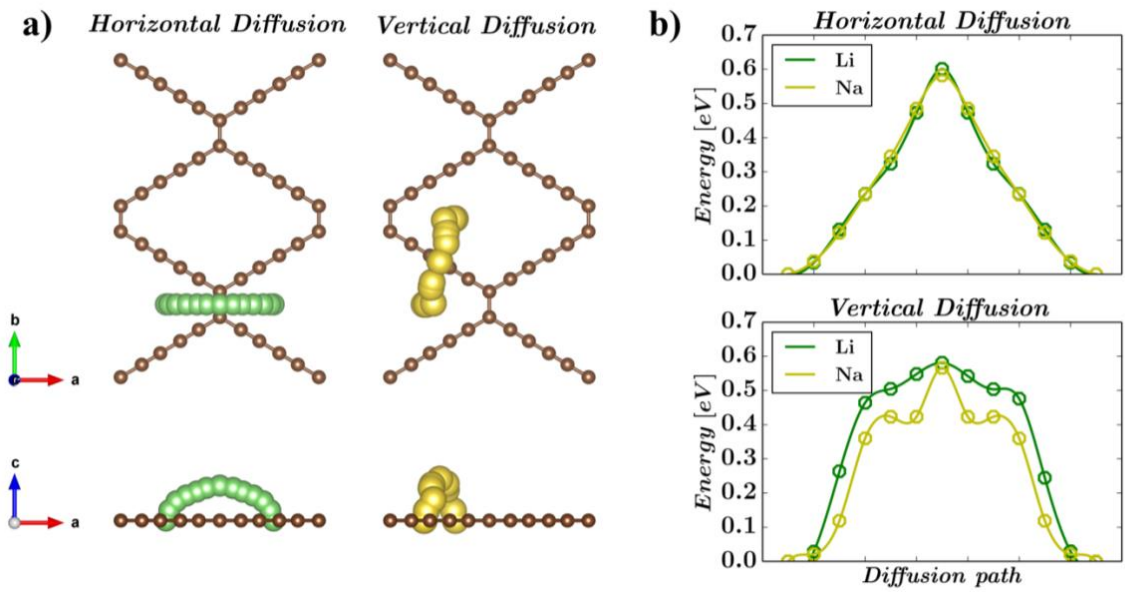

Figure 8. Diffusion of a single Li (green) and Na (yellow) adatom over CEY monolayer: a) Horizontal and vertical pathways; b) Nudged-elastic band method results for the corresponding diffusion pathways.

# Carbon ene-yne graphyne monolayer as an outstanding anode material for Li/Na ion batteries

Meysam Makaremi,[1] Bohayra Mortazavi,[2] and Chandra Veer Singh*,[1,3]

[1]Department of Materials Science and Engineering, University of Toronto, 184 College Street, Suite 140, Toronto, ON M5S 3E4, Canada.
[2]Institute of Structural Mechanics, Bauhaus-Universität Weimar, Marienstr. 15, D-99423 Weimar, Germany.
[3]Department of Mechanical and Industrial Engineering, University of Toronto, 5 King's College Road, Toronto M5S 3G8, Canada.

Table S1. Lattice vectors of the carbon carbon ene-yne (CEY) unitcell.

| Vector | x [Å] | y [Å] | z [Å] |
|---|---|---|---|
| 1 | 11.259 | 0.000 | 0.000 |
| 2 | 0.000 | 9.74 | 0.000 |
| 3 | 0.000 | 0.000 | 20.000 |

Table S2. Atomic positions of the carbon carbon ene-yne (CEY) unitcell.

| Atom | x [Å] | y [Å] | z [Å] |
|---|---|---|---|
| C | 0.000 | 9.713 | 10.000 |
| C | 0.000 | 1.419 | 10.000 |
| C | 1.194 | 2.133 | 10.000 |
| C | 2.246 | 2.783 | 10.000 |
| C | 3.383 | 3.479 | 10.000 |
| C | 4.435 | 4.129 | 10.000 |
| C | 5.629 | 4.842 | 10.000 |
| C | 10.065 | 2.133 | 10.000 |
| C | 9.012 | 2.783 | 10.000 |
| C | 7.876 | 3.479 | 10.000 |
| C | 6.823 | 4.129 | 10.000 |
| C | 5.629 | 6.290 | 10.000 |
| C | 1.194 | 9.000 | 10.000 |
| C | 2.246 | 8.349 | 10.000 |
| C | 3.383 | 7.654 | 10.000 |
| C | 4.435 | 7.003 | 10.000 |
| C | 10.065 | 9.000 | 10.000 |
| C | 9.012 | 8.349 | 10.000 |
| C | 7.876 | 7.654 | 10.000 |
| C | 6.823 | 7.003 | 10.000 |

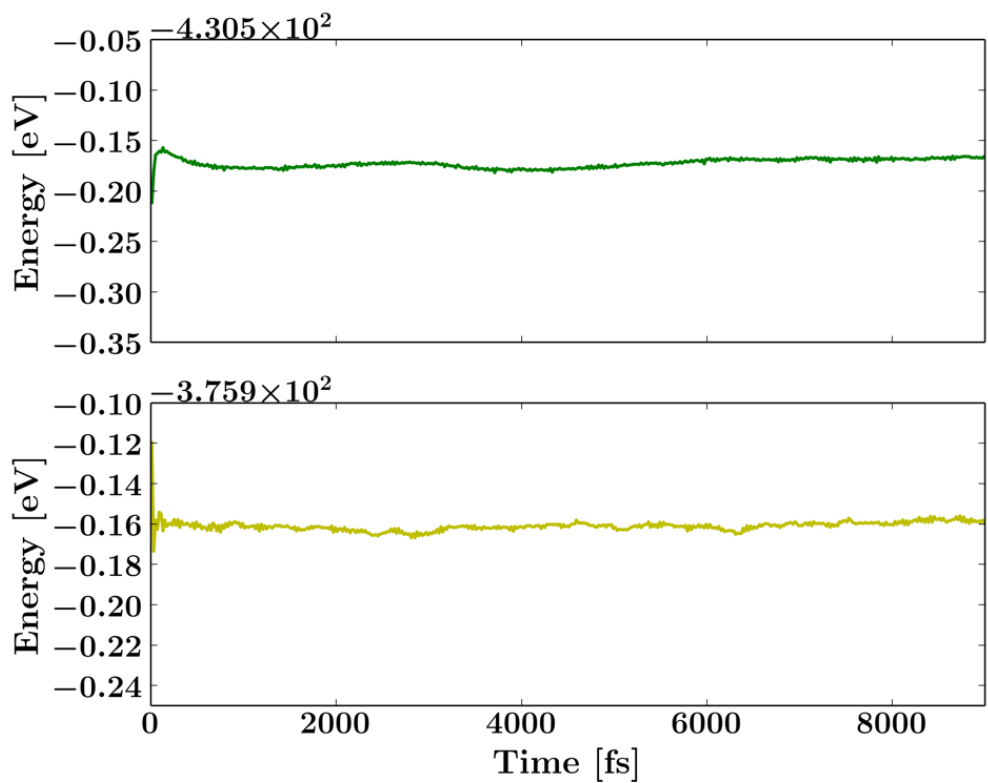

Figure S1. Ab-initio molecular dynamics (AIMD) results for thermal stability. Energy with respect to the simulation time for CEY interacting with Li (green)/Na (yellow) adatoms with the optimal capacity.

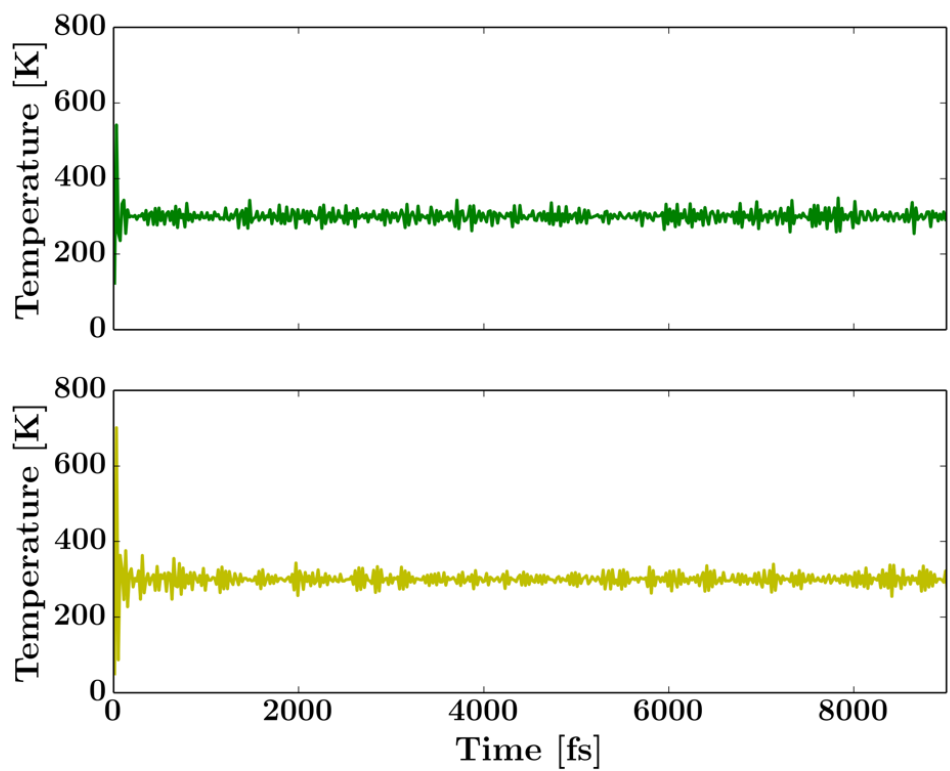

Figure S2. Ab-initio molecular dynamics (AIMD) results for thermal stability. Temperature with respect to the simulation time for CEY interacting with Li (green)/Na (yellow) adatoms with the optimal capacity.